\documentclass[conference]{IEEEtran}
\IEEEoverridecommandlockouts

\usepackage{cite}
\usepackage{amsmath,amssymb,amsfonts}
\usepackage{algorithmic}
\usepackage{graphicx}
\usepackage[utf8]{inputenc}
\usepackage{textcomp}
\usepackage{xcolor}
\usepackage{multirow}
\usepackage{float} 
\usepackage{tabularx}
\usepackage{algorithm}
\usepackage{adjustbox}
\usepackage{soul}
\sethlcolor{yellow}

\usepackage{bm}
\makeatletter
\AtBeginDocument{\DeclareMathVersion{bold}
\SetSymbolFont{operators}{bold}{T1}{times}{b}{n}
\SetMathAlphabet{\mathrm}{bold}{T1}{times}{b}{n}
\SetMathAlphabet{\mathit}{bold}{T1}{times}{b}{it}
\SetMathAlphabet{\mathbf}{bold}{T1}{times}{b}{n}
\SetMathAlphabet{\mathtt}{bold}{OT1}{pcr}{b}{n}
\SetSymbolFont{symbols}{bold}{OMS}{cmsy}{b}{n}
\renewcommand\boldmath{\@nomath\boldmath\mathversion{bold}}}
\makeatother
\def\BibTeX{{\rm B\kern-.05em{\sc i\kern-.025em b}\kern-.08em
    T\kern-.1667em\lower.7ex\hbox{E}\kern-.125emX}}

\begin{document}

\title{Intrusion Detection in IoT Networks Using Hyperdimensional Computing: A Case Study on the NSL-KDD Dataset}

\author{\IEEEauthorblockN{\IEEEauthorrefmark{1} Ghazal Ghajari,\IEEEauthorrefmark{2} Elaheh Ghajari, \IEEEauthorrefmark{1}Hossein Mohammadi, \IEEEauthorrefmark{1} Fathi Amsaad}
\IEEEauthorblockA{\IEEEauthorrefmark{1}Computer Science and Engineering, Wright State University, Ohio, USA }
\IEEEauthorblockA{\IEEEauthorrefmark{2}Computer Science and Engineering, Azad University, Ahvaz, Iran}
Email: \IEEEauthorrefmark{1} \{ghajari.2, mohammadi.5, fathi.amsaad\}@wright.edu
\IEEEauthorrefmark{2} elaheh.ghajari.1@gmail.com
}

\maketitle

\begin{abstract}
The rapid expansion of Internet of Things (IoT) networks has introduced new security challenges, necessitating efficient and reliable methods for intrusion detection. In this study, a detection framework based on hyperdimensional computing (HDC) is proposed to identify and classify network intrusions using the NSL-KDD dataset, a standard benchmark for intrusion detection systems. By leveraging the capabilities of HDC, including high-dimensional representation and efficient computation, the proposed approach effectively distinguishes various attack categories such as DoS, probe, R2L, and U2R, while accurately identifying normal traffic patterns. Comprehensive evaluations demonstrate that the proposed method achieves an accuracy of 99.54\%, significantly outperforming conventional intrusion detection techniques, making it a promising solution for IoT network security. This work emphasizes the critical role of robust and precise intrusion detection in safeguarding IoT systems against evolving cyber threats.
\end{abstract}


\section{Introduction}
As IoT devices become increasingly integrated into everyday life, their security has emerged as a critical concern. These networks, characterized by vast scale and heterogeneity, are particularly vulnerable to cyber threats, such as denial-of-service (DoS) attacks, probing activities, and unauthorized access attempts. Ensuring the security and reliability of IoT systems requires effective methods for detecting and responding to network intrusions.

Intrusion detection plays a pivotal role in IoT security by identifying malicious activities that threaten the integrity, confidentiality, and availability of data within critical systems \cite{shah2023deep, wang2024exploring}. This field can be broadly divided into three approaches: clustering, classification, and anomaly detection. Clustering helps group similar network activities and identify unknown attack types, while classification labels traffic based on predefined categories like DoS, probe, R2L, and U2R. Anomaly detection identifies patterns that deviate from normal behavior, making it particularly effective for detecting zero-day attacks \cite{nguyen2024unknown, ehmer2024network}. Beyond IoT, anomaly detection has critical applications in healthcare, where it helps identify abnormal patterns in medical data, ensuring patient safety \cite{khan2024anomaly, ghajari2024hybrid}.

Many existing intrusion detection systems rely on traditional machine learning techniques, which often struggle with the high-dimensional and dynamic nature of IoT network data \cite{yan2018effective}. These limitations hinder their ability to accurately classify a wide range of attack types while distinguishing malicious activities from normal network behavior. As a result, there is a growing need for solutions that address these challenges with high efficiency and adaptability.

In this work, we introduce a hyperdimensional computing (HDC)-based classification framework for intrusion detection in IoT networks. HDC leverages high-dimensional vector representations to capture complex data patterns, enabling rapid and accurate classification of network traffic. This approach excels at handling high-dimensional data and demonstrates strong generalization capabilities, making it particularly suitable for IoT applications.

This study uses the NSL-KDD dataset, a comprehensive benchmark for network intrusion detection, to evaluate the proposed method. The dataset includes diverse attack categories—DoS, probe, R2L, and U2R—alongside normal traffic, providing a robust platform to assess the model's effectiveness. The results show that the proposed HDC-based framework outperforms traditional techniques in accurately identifying attack types and normal network behavior.

The remainder of this paper is structured as follows: Section 2 explores related work in IoT intrusion detection and hyperdimensional computing. Section 3 details the methodology, including data preprocessing and model design. Section 4 presents experimental results, and Section 5 concludes with insights and future research directions.

\section{Related Work}
Intrusion detection in IoT networks has become a critical area of research due to the increasing complexity and security challenges in these environments. Traditional methods, such as statistical analysis, rule-based approaches, and early machine learning models, struggle to handle the high-dimensional, dynamic, and heterogeneous nature of IoT data. To address these limitations, recent advancements have focused on deep learning techniques like CNNs, RNNs, LSTMs, and GANs, along with ensemble learning strategies, to enhance detection accuracy and identify complex attack patterns \cite{khekare2023optimizing, chemmakha2024towards, hashemitaheri2024optical}. Various studies have explored improvements in IDS by refining machine learning models; for instance, decision trees with optimized feature selection \cite{guezzaz2021reliable} and entropy-weighted KNN \cite{bach2021improvement} have shown enhanced classification performance on the NSL-KDD dataset.

A hybrid approach combining multiple classifiers has also demonstrated promising results in detecting sophisticated attacks. A double-layered hybrid model \cite{wisanwanichthan2021double}, using Naïve Bayes and SVM classifiers for detecting DoS and rare attacks like Remote2Local (R2L) and User2Root (U2R), showed effectiveness on the NSL-KDD dataset. Similarly, another approach \cite{harini2023effective} used a combination of deep neural networks, CNN, LSTM, and XGBoost classifiers to detect minority attacks, also proving effective on the same dataset. Furthermore, adversarial learning-based approaches like FlowGANAnomaly \cite{li2024flowgananomaly} reduce reliance on labeled data by capturing network traffic patterns in an adversarial setting, while optimization-driven methods like SPIDER \cite{udas2022spider} and hybrid IDS frameworks \cite{hussain2022hybrid} leverage dimensionality reduction and evolutionary algorithms to improve intrusion detection performance.

More recent studies have explored improving IDS in specific domains, such as smart grid networks, by integrating feature selection techniques like the African Vulture Optimization Algorithm (AVOA) with deep learning models such as DBN-LSTM \cite{alsirhani2023implementation}. Another study \cite{hanafi2024intrusion} employed an Improved Binary Golden Jackal Optimization (IBGJO) algorithm combined with LSTM to enhance classification performance on benchmark datasets like NSL-KDD and CICIDS2017. Despite these advancements, existing IDS solutions still face challenges, including high false alarm rates, computational complexity, and difficulties in handling imbalanced datasets.

While traditional and deep learning-based intrusion detection models have made significant advancements in accuracy and performance, they still face inherent challenges. Decision tree and KNN-based models \cite{guezzaz2021reliable, bach2021improvement} improve classification accuracy through feature selection but require extensive preprocessing and lack scalability for large-scale IoT networks. Hybrid and deep learning-based methods \cite{wisanwanichthan2021double, harini2023effective} achieve higher accuracy but demand significant computational resources, limiting their applicability in resource-constrained environments. Additionally, adversarial learning techniques like FlowGANAnomaly \cite{li2024flowgananomaly} improve generalization but introduce high training complexity and reduced interpretability.

In contrast, hyperdimensional computing (HDC) offers a lightweight and efficient alternative by encoding feature interactions in a high-dimensional space, eliminating the need for extensive preprocessing and manual feature engineering. Unlike PCA-based models like SPIDER \cite{udas2022spider}, HDC inherently preserves feature integrity while ensuring computational efficiency. However, like other methods, HDC faces challenges related to real-time scalability and adaptability to evolving attack patterns. Future research should focus on integrating HDC with online learning mechanisms and optimizing its implementation for large-scale, dynamic IoT environments. This study builds upon these prior advancements by proposing an HDC-based classification approach for intrusion detection in IoT networks, aiming to enhance detection accuracy while maintaining computational efficiency, particularly in resource-constrained settings.

\section{Methodology}
\begin{figure*}
    \centering
    \includegraphics[width=1\linewidth]{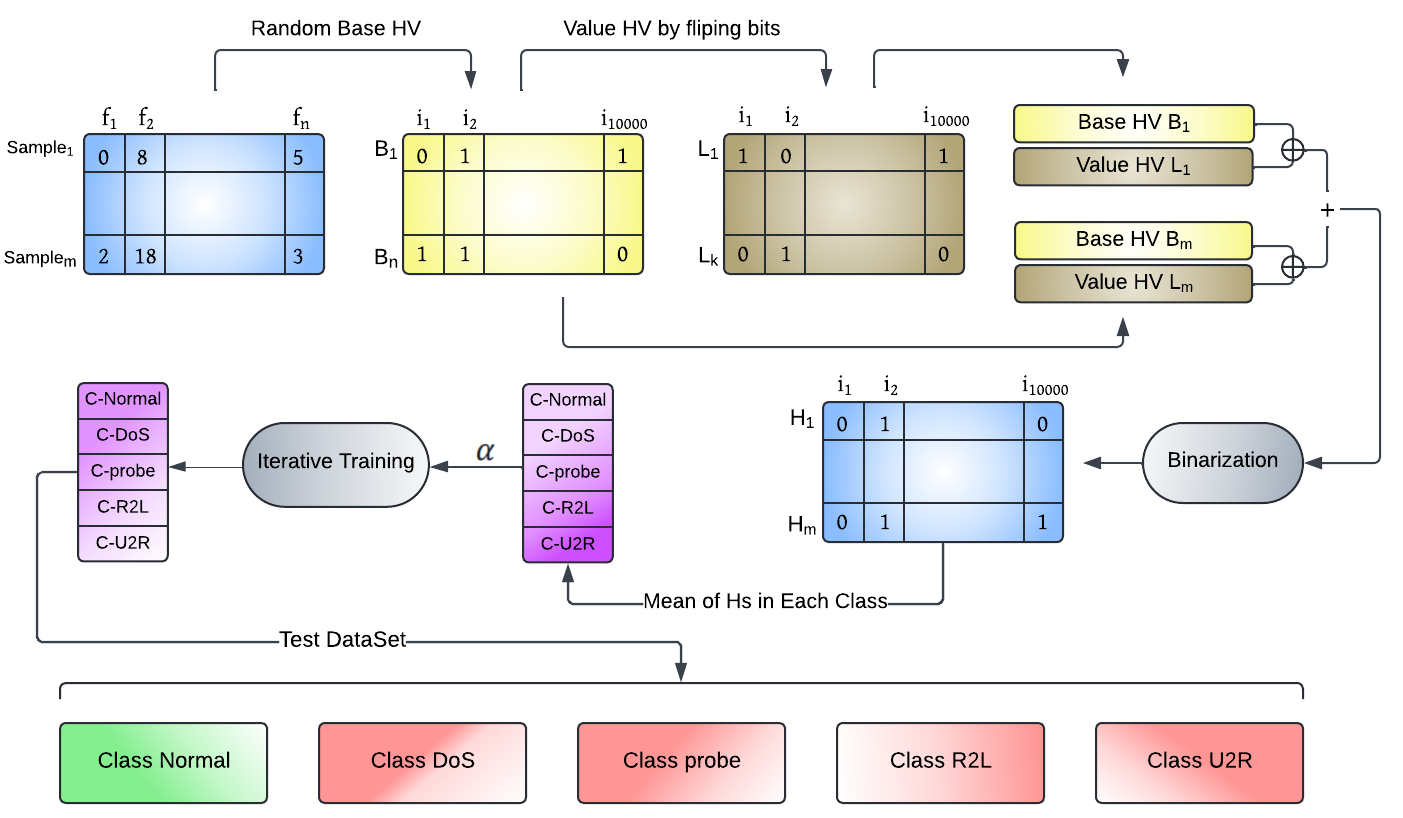}
    \caption{Overview of the proposed hyperdimensional computing (HDC)-based intrusion detection framework. The process begins with feature encoding, where each network sample’s attributes are transformed into high-dimensional binary hypervectors. Value hypervectors are generated by flipping bits systematically to encode feature variations. These base and value hypervectors are combined using element-wise operations to create a unique representation for each sample. The binarization step optimizes computational efficiency by converting continuous values into binary format. During training, class-specific hypervectors are iteratively refined using the mean of encoded samples per class, enhancing classification accuracy. Finally, new test samples are classified based on their similarity to learned class representations, distinguishing between normal and attack categories (DoS, probe, R2L, U2R). This framework effectively balances accuracy and computational efficiency, making it suitable for intrusion detection in IoT networks.}
    \label{fig1}
\end{figure*}

In this section, a high-dimensional encoding approach is proposed to define hypervectors that represent the similarity of different attack classes. These hypervectors are constructed in a manner that ensures the distinctiveness and orthogonality necessary for accurate classification. By training the model using these high-dimensional hypervectors, the system is enabled to effectively detect a wide range of attack types. The following subsections provide a detailed description of the encoding process and the model's architecture. Further technical details are presented in Algorithm 1 and Figure \ref{fig1}, which illustrate the step-by-step approach and the model structure, respectively.

\subsection{High-Dimensional Encoding}
The High-Dimensional (HD) Encoding process consists of multiple systematic stages designed to effectively represent data and facilitate classification:

\subsection*{Step1: Base Hypervectors}
In this approach, each feature is represented by a unique binary hypervector of 10,000 dimensions, denoted as $B_1, B_2, . . .,B_N$    , where $N$ is the total number of features. These hypervectors are generated randomly to maintain orthogonality, as randomly created binary vectors exhibit negligible correlation. This orthogonality ensures the distinct representation of features, preserving their independence and contributing to robust encoding.

\subsection*{Step2: Value Hypervectors}
For each feature, a series of value hypervectors ${L_1, L_2, . . .,L_n}$, each with 10,000 dimensions, is created to encode discrete ranges of feature values. The first hypervector in this set is initialized randomly, while subsequent hypervectors are derived by flipping $\frac{1}{n}$ bits of the previous one. This methodology ensures that the hypervectors representing different bins of feature values are distinguishable while maintaining a structured relationship across bins.

\subsection*{Step3: Encoding Process}
The encoding of a data sample is achieved by combining the base and value hypervectors for all features. The resultant hypervector $H$ is computed using an XOR operation between the base hypervector of a feature and the corresponding value hypervector for its observed range, followed by summation across features:

\[H = B_1 \oplus L_1 + B_2 \oplus L_2 + \dots + B_N \oplus L_N\]

Here, $B_i$ represents the base hypervector of the $i$-th feature, and $L_i$ corresponds to the appropriate value hypervector based on the observed value of that feature. This process creates a distinctive high-dimensional representation for each sample.

\subsection*{Step4: Binarization}
To optimize computational efficiency, the resulting hypervector $H$ is binarized. Each component of $H$ is transformed into either 0 or 1 based on a threshold $T$, which is typically set to half the number of features:

\[H_{\text{bin}}[i] =
\begin{cases} 
1, & \text{if } H[i] > T \\
0, & \text{otherwise}
\end{cases}\]
\subsection*{Step5: Class Representatives}
Class representatives, denoted as $C$, are computed by averaging the binary hypervectors of all samples within a given class. These centroids serve as prototypes for their respective classes:
\[C = \sum_{i=1}^{n} H_{\text{bin}}(i)
\]
Here, $H_{bin}(i)$ refers to the binary hypervector of the $i$-th sample in the class, and $n$ is the number of samples in that class.

\subsection*{Step6: Inference}
For a new data sample, its binary hypervector $H_{bin}(new)$ is compared with the class representatives $C_i$. The class with the highest cosine similarity is assigned to the sample:
\[\text{Cosine Similarity} = \frac{H_{\text{bin}}(\text{new}) \cdot C_i}{\|H_{\text{bin}}(\text{new})\| \|C_i\|}
\]
\subsection*{Step7: Iterative Training}
To improve classification accuracy, iterative updates are applied to the class representatives. For misclassified samples, adjustments are made using the following equations:
\[C_{\text{miss}} = C_{\text{miss}} - \alpha \cdot H_{\text{bin}}(\text{new})\]
\[C_{\text{match}} = C_{\text{match}} + \alpha \cdot H_{\text{bin}}(\text{new})
\]
Here, $\alpha$ is the learning rate. If a sample is correctly classified, no adjustment is needed. Otherwise, the representative of the incorrect class $(C_{miss})$ is updated by subtracting a proportion of the sample hypervector, while the representative of the correct class $(C_{match})$ is updated by adding the same proportion. This iterative adjustment process, typically performed over 50 iterations, refines the classification boundaries for better performance.

\begin{table}
\centering
\begin{tabular}{l}\hline
\textbf{Algorithm 1:} High-Dimensional Encoding Process\vspace{0.1cm}
 \\ \hline
\textbf{Input:} Dataset with $m$ samples and $n$ features\\ 
\textbf{Output:} Binary HyperVector Dataset and Class Representatives\\
$D$: Length of Hypervector (e.g., $D = 10,000$) \\   
$K$: Number of value bins \\   
\textbf{Base Hypervectors:}\\
for each feature $j$ in $n$ features:\\  \hspace{0.2cm}
Create a random binary base hypervector ($B_j$) of length $D$\\
\textbf{Value Hypervectors:}\\
for each feature $j$ in $n$ features:\\  \hspace{0.2cm}
Split the value range \((f_{\text{min}}, f_{\text{max}})\) into $K$ bins\\  \hspace{0.2cm}
Generate $L_1$ randomly as the first value hypervector of length $D$\\
for $i$ in range $(2:K)$:  \hspace{0.2cm}
Create $L_i$ by flipping $D/K$ bits of $L_{i-1}$ \\
Assign $L_i$ to its corresponding bin\\
\textbf{Encoding Process:}\\
for each sample $x$ in $m$ samples:\\  \hspace{0.2cm}
Initialize $H_x$ as a zero hypervector of length $D$\\  \hspace{0.2cm}
for each feature $j$ in the sample:\\  \hspace{0.4cm}
Determine the bin of the feature value and retrieve $\overline{L_j}$\\  \hspace{0.4cm}
Update $H_x = H_x + (B_j \bigoplus \overline{L_j})$\\
\textbf{Binarization:}\\
for each sample $x$ in $m$ samples:\\  \hspace{0.2cm}
for each element $i$ in $H_x$:\\  \hspace{0.4cm}
if $H_x[i] > T$: \hspace{0.2cm}
$H_x[i] = 1$  \hspace{0.2cm}
else:  \hspace{0.2cm}
$H_x[i] = 0$, where $T = n/2$\\
\textbf{Class Representatives:}\\
for each class $c$:  \hspace{0.2cm}
Compute $C_c = \frac{1}{n_c} \sum_{i=1}^{n_c} H_x(i)$, where $n_c$ is \\the number of samples in class $c$\\
\textbf{Inference:}\\
for a new sample $x_{new}$:\\  \hspace{0.2cm}
Compute $H_{bin}(new)$ as above\\  \hspace{0.2cm}
Compare $H_{bin}(new)$ \\  \hspace{0.4cm} with each class representative $C_c$ using cosine similarity):\\  \hspace{0.6cm}
$similarity = \frac{H_{bin}(new) \cdot C_c}{\|H_{bin}(new)\| \|C_c\|}$\\
Assign $x_{new}$ to the class with the highest similarity\\
\textbf{Iterative Training:}\\
for each misclassified sample:\\  \hspace{0.2cm}
Update $C_{miss} = C_{miss} - \alpha \cdot H_{bin}(new)$\\  \hspace{0.2cm}
Update $C_{match} = C_{match} + \alpha \cdot H_{bin}(new)$\\
Repeat updates for 50 iterations\\
\hline
\end{tabular}
\label{alg2}
\end{table}

\subsection{Intrusion Classification with Hypervectors}
The similarity hypervectors defined in the previous stage are utilized to classify the test data into their corresponding intrusion classes. The hypervector for each test sample is compared with the class representatives, which were computed during the training phase by averaging the hypervectors of the samples in each intrusion class. The classification is determined by calculating the cosine similarity between the test sample's hypervector and each class representative. The sample is assigned to the class with the highest similarity score, which indicates the most probable intrusion type. This method allows for efficient categorization of the test data into different intrusion types, ensuring accurate detection and enhancing the system's ability to recognize a variety of intrusions based on the predefined hypervector representations. By utilizing cosine similarity, the most similar class is selected, ensuring that the model performs effective and precise intrusion detection.

\section{Result}
Several studies have evaluated different intrusion detection methods on the NSL-KDD dataset, each proposing novel techniques to improve performance and overcome challenges like execution time and computational complexity.

DTE \cite{guezzaz2021reliable}, which employed min-max normalization and feature selection, achieved 99.42\% accuracy. However, the method was associated with high implementation and execution complexity, making it resource-intensive. EM-KNN \cite{bach2021improvement} achieved an accuracy of 98.83\%. This method did not require preprocessing, but it faced issues with high execution time, which limited its practicality for real-time applications. NB-SVM \cite{wisanwanichthan2021double}, utilizing one-hot encoding, normalization, and PCA, reached an accuracy of 88.97\%, but the performance of this model was highly dependent on the choice of hyperparameters, affecting its flexibility.

The combination of deep learning techniques has also been explored, such as in the WDNN-CNN-LSTM-XGBoost model \cite{harini2023effective}, which achieved 97.94\% accuracy. Despite its effectiveness, the method suffered from high execution times, which made it unsuitable for large-scale or real-time environments. FlowGAN Anomaly \cite{li2024flowgananomaly}, which used normalization, attained an accuracy of 87.47\%. However, it encountered high model complexity, which hindered its deployment and application efficiency. SPIDER \cite{udas2022spider}, utilizing label encoding and PCA for dimensionality reduction, achieved an accuracy of 82.91\%. While effective in some cases, this model faced challenges with imbalanced datasets, which impacted its detection capabilities.

The WOA-ABC-CNN model \cite{hussain2022hybrid}, which used feature selection, achieved 98\% accuracy. Despite this promising result, the method was burdened with high time complexity, posing challenges for real-time applications. AVOA-DBN-LSTM \cite{alsirhani2023implementation}, which used min-max normalization and feature selection, achieved 98.99\% accuracy. However, the model had inadequate test experiments, which limited its broader applicability. IBGJO-LSTM \cite{hanafi2024intrusion}, which also used min-max normalization and feature selection, achieved 98.75\% accuracy but suffered from high time complexity.

In comparison to these methods, the proposed approach using hyperdimensional computing-based classification achieved a remarkable 99.5\% accuracy on the NSL-KDD dataset, as shown in Table \ref{tab1}. This method not only surpasses the accuracy of several existing techniques but also addresses key challenges, such as high execution time and complexity, that were observed in prior studies. the proposed model is able to efficiently handle high-dimensional data, offering an improved solution for intrusion detection in IoT networks.


\begin{table}
\centering
\caption{Accuracy Comparison of Techniques on NSL-KDD Dataset}
\tiny
\resizebox{\columnwidth}{!}{%
\begin{tabular}{l|c}
\hline
\textbf{Technique} & \textbf{Accuracy} \\ \hline
proposed Method (HDC-based) &~~~~ \textbf{99.54\%} ~~~~\\ \hline
DTE \cite{guezzaz2021reliable} & ~~~~ 99.42\% ~~~~\\ \hline
EM-KNN \cite{bach2021improvement} & ~~~~ 98.83\% ~~~~\\ \hline
NB-SVM \cite{wisanwanichthan2021double} & ~~~~ 88.97\% ~~~~\\ \hline
WDNN-CNN-LSTM-XGBoost \cite{harini2023effective}~~~"
~  & ~~~~ 97.94\% ~~~~\\ \hline
FlowGAN Anomaly \cite{li2024flowgananomaly} & ~~~~ 87.47\% ~~~~\\ \hline
SPIDER \cite{udas2022spider} & ~~~~ 82.91\% ~~~~\\ \hline
WOA-ABC-CNN \cite{hussain2022hybrid} & ~~~~ 98.00\% ~~~~\\ \hline
AVOA-DBN-LSTM \cite{alsirhani2023implementation} & ~~~~ 98.99\% ~~~~\\ \hline
IBGJO-LSTM \cite{hanafi2024intrusion} & ~~~~ 98.75\% ~~~~\\ \hline
\end{tabular}
}
\label{tab1}
\end{table}

\section{Conclusion and Future Work}
In this paper, a new approach to intrusion detection in IoT networks utilizing a hyperdimensional computing (HDC)-based classification model is introduced. By exploiting HDC’s ability to effectively represent high-dimensional data and perform fast classification, its capability to detect various attack types, such as DoS, probe, R2L, and U2R, in the NSL-KDD dataset is demonstrated. The experiments showed that the HDC-based model outperforms traditional machine learning techniques in terms of accuracy, making it a promising solution for enhancing IoT network security.

Although the results are promising, there are several areas for further research. Firstly, while the focus was on the NSL-KDD dataset, testing the model on other real-world IoT datasets would be beneficial to evaluate its generalization capabilities. Furthermore, although HDC excels at handling high-dimensional data, improving its computational efficiency for large-scale IoT networks with limited resources remains an important challenge. Future work could also investigate the integration of HDC with other anomaly detection techniques, such as deep learning or ensemble methods, to enhance detection performance.

Another avenue for future research could involve extending the model to handle sophisticated attacks, such as zero-day and advanced persistent threats, and incorporating real-time detection and response for better performance in dynamic, resource-constrained IoT environments.

Overall, this research demonstrates the potential of hyperdimensional computing as an effective tool for intrusion detection in IoT networks. By addressing the challenges associated with high-dimensional data and improving classification accuracy, this approach can contribute to developing more secure and resilient IoT systems in the future.

\ifCLASSOPTIONcaptionsoff
  \newpage
\fi
\bibliographystyle{IEEEtran}

\end{document}